\documentclass{article}[12pt]
\usepackage{amsmath}
\usepackage{latexsym}
\usepackage{float}
\usepackage{amssymb}
\usepackage{graphicx}
\usepackage{epsfig}

\begin{document}

\title{Reply to the Comment by M. Biskup, L. Chayes and R.
Kotecky}

\author{K. Binder \\
Institut f\"ur Physik, Johannes Gutenberg Universit\"at Mainz,
\\ Staudinger Weg 7, 55099 Mainz, Germany}

\maketitle

\abstract{In their comment to the paper ``Theory of the
evaporation/condensation transition of equilibrium droplets in
finite volumes'' [Physica A\underline{319}, 99 (2003)], Biskup et
al. claim that in finite systems at fixed density ``the physical
significance of the conjugate thermodynamic variable is of dubious
value''. This claim is critically discussed.}\\



\medskip

While in the thermodynamic limit the various ensembles of
statistical thermodynamics are trivially related to each other by
Legendre transformations, in finite systems nontrivial differences
of the physical properties in different ensembles appear. This
fact, emphasized in the preceding comment \cite{1}, of course, is
very well known \cite{2}, and it is emphasized in the literature
on computer simulations \cite{3,4,5} in particular, since computer
simulations always have to deal with systems of finite size only.

For instance, considering an Ising ferromagnet for temperatures
below the critical temperature $T_c$, as a function of magnetic
field $H$, the conjugate variable, the magnetization $m$, is a
quantity that exhibits nontrivial fluctuations, but its average
$\langle m \rangle $ increases monotonically from negative to
positive values as $H$ is increased from negative to positive
values. Thus $\langle m \rangle = - \partial f(T,H,L)/\partial H$
is a smooth regular function for all finite linear dimensions $L$,
in the d-dimensional $L^d$ geometry with periodic boundary
conditions, and its derivative $\chi = (\partial \langle m \rangle
/\partial H)_T$ is everywhere nonnegative and finite. When $L
\rightarrow \infty , M(H)$ develops the jump singularity at $H=0$
from $-m_{coex}$ to $+m_{coex}, \; m_{coex}$ being the value of
the spontaneous magnetization. The details of this behavior have
been elucidated in the literature and are well understood
\cite{6,7,8}.

The behavior of the finite Ising magnet where the magnetization
$m$ is held fixed as an independent variable is very different
\cite{9}. Then the conjugate intensive variable $H$ is a quantity
that exhibits nontrivial fluctuations, and its average $\langle H
\rangle = \partial g(T,m,L)/\partial m$ is a smooth regular
function for all finite $L$, but for $T <T_c$ exhibits a loop as
function of $m$, rather than increasing monotonically with
increasing $m$. The singular behavior that occurs in the
thermodynamic limit due to the presence of this loop was discussed
in \cite{9}. Of course, if one keeps $m$ fixed and takes $L
\rightarrow \infty $, then the isotherm develops a constant
horizontal part $H=0$ from $-m_{coex}$ to $+m_{coex}$ (This is
trivially related to the corresponding result in the $(H,T)$
ensemble via the Legendre transformation, of course). The
different singular behavior of $H$ vs. $m$ discussed in \cite{9}
only follows in a special limit where one takes the limits
$L\rightarrow \infty $ and $m \rightarrow -m_{coex}$ (from the
side where $m >-m_{coex}$) together, such that $\delta
\tilde{m}=(m+m_{coex})L^{d/(d+1)}= const$. Since at finite $L$ for
$m$ near $-m_{coex}$ some (rounded) remnants of the transition
present in this limit are still observed in simulations (the first
observation was actually made in \cite{10}), this transition and
its various signatures are of interest for simulations, motivating
the discussion presented in \cite{9}. In fact, in the equivalent
lattice gas interpretation of the Ising model this transition gets
the meaning of an evaporation transition of a droplet in a finite
volume when it reaches a certain minimum size. The existence of
this transition, first suggested in \cite{11} (Eqs. 26,~27), of
that reference) was recently established rigorously \cite{12}.

\begin{figure}[h]
\begin{center}
\epsfig{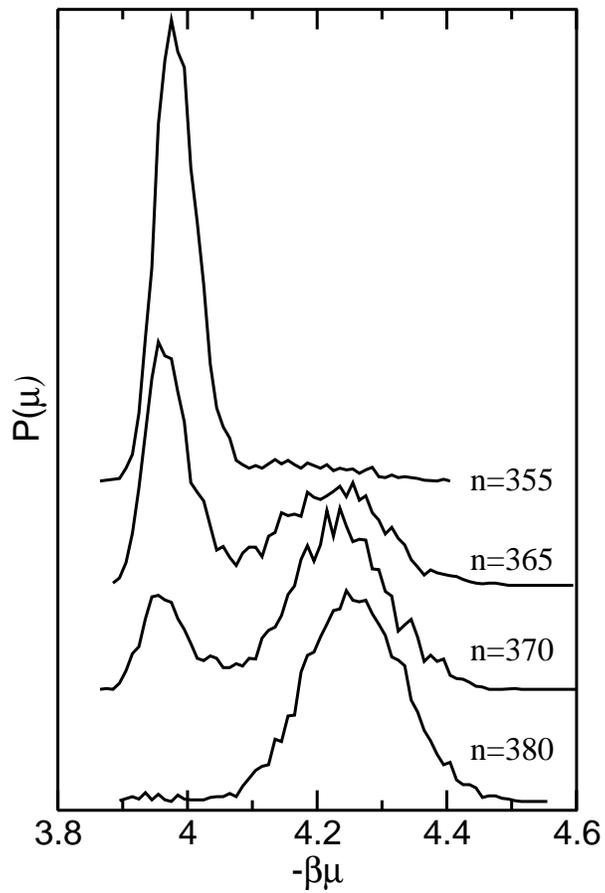}
\caption{\label{fig1}
Distribution $p(\mu)$ of the
chemical potential $\mu$ for a Lennard-Jones fluid at $T=0.68
\varepsilon/k_B$ and several choices of $N$ as indicated, for a
cubic box of volume ($22.5 \sigma)^3$, $\sigma$ being the range
parameter of the Lennard-Jones potential and $\varepsilon$ its
strength. From MacDowell et al. \cite{15}.
}
\end{center}
\end{figure}

Of course, there is nothing wrong to consider the intensive
variable conjugate to a fixed density (magnetization density or
particle number density, etc.) in a finite system. Considering the
fluctuations of the temperature in a finite system in the
microcanonical NVE-ensemble is standard textbook wisdom \cite{13}.
Similarly, the chemical potential $\mu$ in the NVT ensemble for
finite volumes has always been of great interest in the context of
computer simulations \cite{3,4,5} and the Widom particle insertion
method \cite{14} has been especially devised for the sampling of
the chemical potential in the NVT ensemble. If the statement of
Biskup et al. \cite{1} that ``in such systems the meaning of a
conjugate variable is rather murky'' were correct a large body of
well-established work would become obsolete. Even if the chemical
potential in a finite system has an ``at best secondary meaning''
\cite{1} in the context of the rigorous derivations, it is a well
accessible and useful quantity for the simulations, and it gives
useful information for the phenomenon at hand. As an example,
Fig.~\ref{fig1} presents recent data \cite{15} on the distribution
of the chemical potential in a three-dimensional Lennard-Jones
fluid for $T \approx 0.68 T_c$ in a box of size $L = 22.5 \sigma$
($\sigma$ being the range of the Lennard-Jones potential) for
several values of the particle number $N$ near the droplet
evaporation/condensation transition. One can see that the
distribution of $\mu$ changes from a single peak distribution for
$N=355$ representing the strongly super-saturated gas without a
droplet to a two peak distribution near $N \approx 370$, where
part of the time a droplet is present in the system, and part of
the time of the sampling it is not present, while for $N=380$ only
the second peak remains, describing the state where the droplet
always coexists with the surrounding, less strongly
supersaturated, gas. This is exactly the behavior suggested on the
basis of the considerations described in \cite{9}.

In conclusion, we do not agree with the basic claim of the
preceding comment, namely that in finite systems at fixed density
``the physical significance of the conjugate thermodynamic
variable is of dubious value'', but we also would like to
emphasize that in science the question whether something is
``valuable'' or not is basically subjective, and the real question
about scientific results is whether they are right or wrong. Thus,
we invite the reader to form his own opinion about this subject.

\underline{Acknowledgements}: I am grateful to L. G. MacDowell, P.
Virnau and M. M\"uller for a fruitful collaboration which resulted
in Ref.~\cite{15}.

\newpage

\newpage

\begin{thebibliography}{99}

\bibitem{1}
M. Biskup, L. Chayes and R. Kotecky, preceding comment.

\bibitem{2}
T. L. Hill, {\it Thermodynamics of small systems} (Benjamin, New
York, 1963)

\bibitem{3}
K. Binder (ed.) {\it Monte Carlo Methods in Condensed Matter
Physics}, Springer, Berlin (1992).

\bibitem{4}
D. Frenkel and B. Smit: {\it Understanding Molecular Simulation:
>From Algorithms to Applications}. Academic, San Diego (1996)

\bibitem{5}
D. P. Landau and K. Binder, {\it A Guide to Monte Carlo
Simulations in Statistical Physics}, Cambridge University Press,
Cambridge (2000)

\bibitem{6}
V. Privman and M. E. Fisher, J. Stat. Phys. \underline{33}, 285
(1983)

\bibitem{7}
K. Binder and D. P. Landau, Phys. Rev. B\underline{30}, 1477
(1984)

\bibitem{8}
C. Borgs and R. Kotecky, J. Stat. Phys. \underline{61}, 79 (1990)

\bibitem{9}
K. Binder, Physica A\underline{313}, 99 (2003)

\bibitem{10}
H. Furukawa and K. Binder, Phys. Rev. A\underline{26}, 556 (1982)

\bibitem{11}
K. Binder and M. H. Kalos, J. Stat. Phys. \underline{22}, 363
(1980)

\bibitem{12}
M. Biskup, L. Chayes and R. Kotecky, Europhys. Lett.
\underline{60}, 21 (2002)

\bibitem{13}
L. D. Landau and M. E. Lifshitz, {\it Statistical Physics},
Pergamon Press, Oxford (1958)

\bibitem{14}
B. Widom, J. Chem. Phys. \underline{39}, 2808 (1963)

\bibitem{15}
L. G. MacDowell, P. Virnau, M. M\"uller, and K. Binder
(unpublished)

\end{thebibliography}
\end{document}